\documentstyle[12pt]{article}
\newcommand{\beq}{\begin{equation}}
\newcommand{\eeq}{\end{equation}}
\newcommand{\beeq}{\begin{eqnarray}}
\newcommand{\eeeq}{\end{eqnarray}}

\begin{document}

\begin{center}

{\bf The interplay between flattening and  damping \\
of  single particle spectra in strongly correlated
Fermi systems} \\
\bigskip
V.~A.~Khodel and M.~V.~Zverev \\
{\small Kurchatov Institute of Atomic Energy, Moscow, 123182, Russia}\\
\end{center}
\begin{abstract}
The self-consistent theory of the fermion condensation, a specific
phase transition which results in
a rearrangement of the single particle degrees of freedom in
strongly correlated Fermi systems is developed. Beyond the phase
transition point, the single particle spectra are shown to be flat.
The interplay between the flattening and the damping of the single
particle spectra at $T\to 0$ is investigated.
The width $\gamma(\varepsilon)$ of the single particle states
is found to grow up linearly with $\varepsilon$
over a wide range of energy  as in a marginal Fermi liquid.
Our results gain insight into the success  of
the phenomenological theory of the normal states
of high-temperature superconductors by Varma et al.

\end{abstract}

\vskip 0.5 cm
\begin{centerline}
{\bf 1. Introduction}
\end{centerline}

\vskip 0.3 cm
In the article \cite{vkvs} within  the Landau-Migdal (LM)
quasiparticle pattern, a new phase transition in strongly correlated Fermi
systems, the so called fermion condensation, has been uncovered.
Its salient feature is in the appearance of the
fermion condensate (FC), a group of degenerate
states whose energy at $T=0$ coincides with the chemical potential
$\mu$. Owing to the degeneracy, the quasiparticle occupation
numbers $n({\bf p},T=0)$ are no longer 1 or 0. They are determined
by a variational condition \cite {vkvs,physrep,noz}
\beq
\delta E_0(n)/\delta n({\bf p})=\mu, \qquad {\bf p}\in \Omega ,
\label{var}
\eeq
where $E_0(n({\bf p}))$ stands for the ground state energy,
while $\Omega$ denotes the FC region whose boundaries are
determined by eq.~(\ref{var}) itself.
According to Landau, the l.h.s.~of (\ref{var})
is nothing but the quasiparticle energy $\varepsilon({\bf p})$
and therefore this equation implies smearing
the Fermi surface --- its metamorphosis it into a volume
in  the three-dimensional systems or into a surface in the
two-dimensional ones. In anisotropic systems,
say, in  crystals, the FC may occupy only part of the old Fermi
surface, e.g.~a region, adjacent to the van Hove
singularities \cite {vol,clark}.

At $T>0$ the quasiparticle distribution  $n({\bf p},T)$ is
found with the help of  the usual Fermi-liquid-theory
variational equation $\delta F/\delta n({\bf p})=0$ \cite{noz}
where $F=E_0-\mu N-TS$ is the free energy.
The solution is written in the standard Fermi-Dirac form
\beq
n({\bf p},T)=\left[1+\exp\left({\xi({\bf p};n)\over T}\right)\right]^{-1} ,
\label{disq}
\eeq
but the energy
$\xi({\bf p})=\varepsilon({\bf p})-\mu$ depends on the distribution
$n$ itself. For both equations (\ref{var}), (\ref{disq}) to be consistent,
the spectrum $\xi({\bf p})$ should exhibit a strong
$T$-dependence in the FC region
\beq
\xi_{FC}({\bf p},T)\simeq T\ln {1{-}n_0({\bf p})\over n_0({\bf p})},
\qquad {\bf p}\in \Omega ,
\label{spc}
\eeq
where $n_0({\bf p})$ is the solution of eq.~(\ref{var}). We see that the
plateau  in the spectrum $\xi({\bf p})$ existing at $T=0$ is converted
into an inclined plane with the slope proportional to $T$.
Such a significant liability of single particle characteristics with
respect to $T$ is a calling card of  fermion condensation.

The flattening of the single particle spectra $\xi({\bf p})$ has been
experimentally observed in the ARPES study  of electron systems of
some high-$T_c$ superconductors \cite {camp,des,shen} but observable
single particle peaks have much larger widths $\gamma$ than those 
calculated employing Fermi liquid theory. As a result, the applicability 
of the quasiparticle formalism  to such systems has been put in question
and alternative explanations have been proposed \cite {schrief,norm}.

As known in Landau theory,
the single particle spectrum $\xi(p)$ is described by the only parameter ---
the effective mass $M^*$, characterizing the slope
$(d\xi(p)/dp)_F\sim (M^*)^{-1}$ of the spectrum at the Fermi surface.
Usually the effective mass is treated as a $T$-independent quantity.
But, as seen from (\ref{spc}), beyond the FC phase transition point,
the slope $d\xi_{FC}/dp$ is just proportional to $T$, i.e. for the FC states
$M^*\sim T^{-1}$. This  peculiarity  of the fermion condensation
alters the Landau theory predictions even if the quasiparticle picture
survives in the systems with the FC. However, the ratio 
$\gamma_{FC}(T)/\xi_{FC}(p,T)$  calculated within the Landau approach
\cite {vrshag,ind,schuck} does not vanish at
$T\to 0$ in contrast to fundamentals of the Fermi liquid theory.
In other words, at finite $T$, the theory of the fermion condensation
based on the LM quasiparticle concept suffers from inconsistency.

On the other hand, at $T=0$, the quasiparticle picture is known
to hold as long as the critical index $\nu$ characterizing behavior of
the imaginary part of the mass operator
${\rm Im}\,\Sigma(p,\varepsilon) \sim \varepsilon^{\nu}$, exceeds 1
 (here $\varepsilon$ is measured from the chemical potential $\mu$).
This conclusion stems from
the formula for the renormalization constant $z$ given by
\beq
z=\biggl[1-{\rm Re}\,\biggl( {\partial \Sigma(p_F,\varepsilon)
\over\partial\varepsilon}
\biggr)_{\varepsilon=0}\,\biggr]^{-1} .
\label{fact}
\eeq
Straightforward calculations with taking into account the well known
dispersion relation \cite {trio,mig}
\beq
{\rm Re}\,\biggl(\Sigma( p,\varepsilon)-\Sigma( p,0)\biggr)=
 {1\over \pi}P\int{ {\rm Im}\,\Sigma(p,\varepsilon_1)\,
{\mbox{sgn}}\,\varepsilon_1\over
\varepsilon_1(\varepsilon_1-\varepsilon)}\,d\varepsilon_1,
\label{dis}
\eeq
demonstrate that the factor $z$ vanishes provided $\nu\leq 1$. Thus
if $\nu\leq 1$ the conventional quasiparticle picture is destroyed.
We shall see that this just occurs beyond the point of the fermion
condensation.

The interplay between damping and flattening beyond this point
is quite subtle. Indeed, let the solution with the FC, i.e.~with
the plateau in $\xi({\bf p})$ at $T=0$, exist neglecting damping
effects. When damping effects get involved the plateau in
$\xi({\bf p})$ should somehow change. If the flat portion  of
$\xi({\bf p})$ were completely destroyed then  the spectrum $\xi({\bf p})$
could be characterized by the single constant $M^*$ that results in 
reviving  Fermi liquid theory and the FC solution, as well.
The way out is that the flat portion
in the single particle spectrum  survives but flattening and damping
should somehow counterbalance each other.

In this article, we develop the self-consistent theory of the
fermion condensation proceeding from the standard many-body-theory
approach and bypassing the variational condition (\ref{var})
which is not adapted to take damping effects into account. The first step
is to derive a general relation between the spectrum $\xi({\bf p})$ and
the distribution $n({\bf p})$ assuming
the LM quasiparticle pattern to be valid. The second step
is to verify that this relation considered as an
equation has a class of peculiar solutions for which $\xi(n({\bf p}))=0$ in
a whole region $\Omega$ of the momentum space rather than at several 
isolated points. These are just the
FC solutions. However, as we shall see, the LM quasiparticle picture
is nothing more than a reasonable approximation for the treatment of
the systems with the FC. These systems do possess features inherent in
marginal Fermi liquids for which the quasiparticle weight in the
single particle state adjacent to the Fermi surface approaches zero.
 The next step is to extend the Landau formalism to marginal
Fermi liquids including the derivation of the relation between the
single particle spectrum and the distribution function. Then one needs
to calculate irregular contributions to the single particle
mass operator beyond the FC phase transition point
and evaluate how strong the plateau in the spectrum of the single
particle excitations is inclined due to the damping.

\vskip 0.5 cm
\begin{centerline}
{\bf 2. The fermion condensation in the quasiparticle approximation}
\end{centerline}
\vskip 0.3 cm

We already mentioned above that the systems with the FC possess the
properties inherent in marginal Fermi liquids. In particular,
the single particle mass operator $\Sigma({\bf p},\varepsilon)$
contains irregular terms precluding expansion
in a Taylor series in the  energy $\varepsilon$.
However, these terms  rapidly die out if 
$\varepsilon$ exceeds a characteristic FC energy $\varepsilon_{FC}$.
Therefore the quasiparticle picture seems to be a reasonable
first approximation for studying the energy dependence of quantities
of our interest.

In the system with the FC, the quasiparticle Green function $G^q$
is found by retaining leading terms in the expansion of the inverse
Green function $G^{-1}$ in the limit $\varepsilon\to 0$.
This expansion reads as
\beq
G^{-1}({\bf p},\varepsilon\to 0)=\varepsilon-\xi^0_p-
\Sigma({\bf p},\varepsilon{=}0)-\varepsilon\,
\biggl({\partial\Sigma(p_c,\varepsilon) \over
\partial\varepsilon}\biggr)_0,
\eeq
where $\xi^0_p=p^2/2M-\mu$. Here we neglected the variation of the
derivative
$(\partial\Sigma(p_c,\varepsilon)/\partial\varepsilon)_0$
in the FC region and calculated this derivative at a fixed point 
$p_c\in\Omega$.

The pole term $G^q$ has the accustomed form
$$
G^q({\bf p},\varepsilon)=\biggl(\varepsilon-\xi({\bf p})\biggr)^{-1}.
$$
However, the quasiparticle spectrum $\xi({\bf p})$ near the Fermi
surface is given by a different formula
\beq
\xi({\bf p})= z\,(\xi^0_p+
\Sigma({\bf p},\varepsilon{=}0)),
\eeq
than in the Landau theory of normal Fermi liquid since the momentum
expansion with retaining terms of the type
$(p^2{-}p^2_F)(\partial\Sigma/\partial p^2)_F$
makes no sense in the system with the FC.

The quasiparticle momentum distribution is evaluated as usual
\beq
n(\xi({\bf p}))={\rm Im}
\int G^q({\bf p},\varepsilon){d\varepsilon\over 2\pi i} ,
\label{mom}
\eeq
yielding the step function $\theta(\xi({\bf p}))$.

Another relation between $\xi$ and $n$ is found proceeding from
the Landau-Pitaevskii identity \cite {trio,mig}
\beq
{\partial G^{-1} ({\bf p},\varepsilon)\over \partial p_{\alpha}}=
{p_{\alpha}\over M}+\int U ({\bf p},\varepsilon,{\bf p}_1,\varepsilon_1)
\,G({\bf p}_1,\varepsilon_1)\,
{\partial G^{-1} ({\bf p}_1,\varepsilon_1)\over \partial p_{1\alpha}}
\,G({\bf p}_1,\varepsilon_1)\,{d^4p_1 \over (2\pi)^4i}.
\label{id}
\eeq
Here $U$ stands for the block of diagrams irreducible in the p-h channel.

In doing the renormalization of eq.~(\ref{id}) a little modification
of the standard decomposition \cite {trio,mig}
of the product $GG$ into a regular part $B$ and a singular one $A$
is necessary. It refers to the singular term $A$. Now it is defined as follows
\beq
A({\bf p},{\bf q},\varepsilon,\omega)=z^2\delta(\varepsilon)\int
G^q({\bf p},\varepsilon)\,G^q({\bf p}{+}{\bf q},\varepsilon{+}\omega)
\,{d\varepsilon\over 2\pi i}.
\label{dec}
\eeq
Then
$$
\int{\partial G^{-1} ({\bf p},\varepsilon)\over
\partial p_{\alpha}}A({\bf p},\varepsilon){d\varepsilon\over 2\pi i}=
z^2 \int G^q({\bf p},\varepsilon)\,
{\partial G^{-1} ({\bf p},\varepsilon{=}0)\over
\partial p_{\alpha}}\, G^q({\bf p},\varepsilon)
{d\varepsilon\over 2\pi i}=z{\partial n({\bf p})
\over \partial p_{\alpha}}.
$$
In obtaining this result, we have used relation
$$
{\partial G^{-1} ({\bf p},\varepsilon{=}0)\over \partial p_{\alpha}}=
z^{-1}{\partial (G^q({\bf p},\varepsilon))^{-1}\over \partial p_{\alpha}}.
$$
As a result, after conventional transformations, we arrive at the
well known relation \cite {trio,mig}
\beq
{\partial \xi({\bf p},n)\over \partial p_{\alpha}}={p_{\alpha}\over M}+
\int {\cal F}({\bf p},{\bf p}_1;n)\,{\partial n({\bf p}_1)
\over \partial p_{1\alpha}}\,d\tau_1 .
\label{lp}
\eeq

The phenomenological quantity ${\cal F}\sim (1-BU)^{-1}U$, being 
$\omega$-independent,
plays the role of an effective interaction potential between particles
in medium. Assuming this function to be known including possible dependence
of ${\cal F}$ on the distribution $n(\xi)$ itself, eq.~(\ref{lp}) can be 
integrated and conveniently rewritten as
\beq
\xi({\bf p})={p^2\over 2M}+\int H({\bf p},{\bf p}_1;n)\,n(\xi({\bf p}_1))\,
d\tau_1,
\label{main}
\eeq
the effective Hartree-like term $H({\bf p},{\bf p}_1)$ being related to 
the effective interaction ${\cal F}$ by the formula 
$$
{\cal F}=H+ n{\delta H\over \delta n}.
$$
As mentioned above, the momentum distribution function $n(\xi)$
 is simply $\theta(\xi)$ so that eq.~(\ref{main})
can be treated as an equation for finding the spectrum $\xi({\bf p})$.
 Usually ${\cal F}$ and, hence, $H$ are  nearly
momentum independent functions. Then 
eq.~(\ref{main}) has the single solution $\xi(p)$ for any $ p$,
with $d\xi/dp>0$, i.e.~there is one-to-one correspondence between
$\xi$ and $p$.

The situation drastically changes in strongly correlated systems where
momentum dependent components of the effective interaction can
be strong. The case of homogeneous matter where particle interact by
means of limited long range effective forces, i.e.
$H({\bf p},{\bf p}_1)=V\delta({\bf p}-{\bf p}_1)$, analyzed in
\cite {noz} is especially impressive. In this case, relation
between $\xi$ and the distribution $n$ becomes algebraic:
\beq
\xi= \xi^0_p+Vn(\xi).
\label{rel}
\eeq
This formula considered as an equation for finding $\xi(p)$ is
rewritten in the form
\beq
L(\xi,p)\equiv {\xi-\xi^0_p\over V}=n(\xi) ,
\label{eqnoz}
\eeq
where $n(\xi)=\theta(\xi)$.
This equation has a simple graphical solution (see Fig.~1) that
furnishes a transparent explanation of how the FC solution of
eq.~(\ref{main}) arises in more complicated cases. Indeed,
the l.h.s.~$ L(\xi,p)$ of eq.~(\ref{eqnoz}) depending explicitly
on the momentum $p$ as a parameter is depicted by a set of straight
lines with the slope $\sim 1/V$. On the other hand, the r.h.s.~$n(\xi)$,
made up of two horizontal lines with the kink at $\xi=0$, is
independent of $p$. Crossing points of the set $L(\xi,p)$ with the
horizontal pieces of $n(\xi)$ are of no interest while in the entire 
interval $p_i<p<p_f$ satisfying the inequality
\beq
  -V<\xi^0_p<0 ,
\label{cond}
\eeq
the straight lines cross the vertical line $\xi=0$. Thus all particles 
with the momenta $p$ varying in this interval turn out to have the same 
energy $\xi(p)=0$, i.e. they belong to the FC region.
It is worth noting that the FC solution is stable provided $V>0$. 

The graphical procedure is easily generalized to the case of nonzero 
temperatures where the distribution function $n({\bf p})$ is given by 
eq.~(2). This model is also  applied to anisotropic systems \cite{vol} 
where the energy $\xi^0_{{\bf p}}$ becomes angle-dependent that results 
in the anisotropy of the FC occupation. So, in a two-dimensional
tight-binding model with the single particle spectrum given by
\beq
\xi^0_{{\bf p}}=-\varepsilon_0(\cos p_x+\cos p_y+t\cos p_x\cos p_y)
-\mu,
\label{ani}
\eeq
and $t>0$, the condition (\ref{cond}) of the model for particles becoming 
trapped into the FC reads as
\beq
0<\cos p_x+\cos p_y+t\cos p_x\cos p_y
+{\mu\over\varepsilon_0}<{V\over\varepsilon_0}.
\label{ineq}
\eeq
It is worth noting that often components of the effective interaction 
between electrons in solids have quite narrow peaks. So, the spin-spin 
component considered in the first quadrant of the quadratic two-dimensional 
Brillouin zone has the peak close to the  point ${\bf Q}=(\pi,\pi)$ 
associated with antiferromagnetism \cite{kampf, mason}. 
The resonance part of this interaction can be crudely written in the form
\beq
S({\bf p}_1,{\bf p}_2,{\bf q})=
-\lambda 
\,\vec\sigma_1\vec\sigma_2
\,\phi({\bf p}_1)\,\phi({\bf p}_2)\,\delta({\bf q}{-}{\bf Q}).
\label{exch}
\eeq
This is an exchange spin fluctuation term. Inserting it into the RPA 
formula for the mass operator $\Sigma$ an essentially momentum dependent 
contribution 
\beq 
\Sigma({\bf p})={\lambda\over 2}\,\phi({\bf p})\,\phi({\bf p}{-}{\bf Q})\,
 n({\bf p}{-}{\bf Q})
\label{sigex}
\eeq 
arises (see Fig.~2). Assuming other contributions to be reduced to 
a renormalization of the chemical potential a system of two equations
\begin{eqnarray}
\xi_1&=&\xi^0({\bf p})+{\lambda\over 2}\,\phi({\bf p})\,
                 \phi({\bf p}{-}{\bf Q})\,n(\xi_2), \nonumber \\
\xi_2&=&\xi^0({\bf p}{-}{\bf Q})+{\lambda\over 2}\,\phi({\bf p})\,
                 \phi({\bf p}{-}{\bf Q})\,n(\xi_1)
\end{eqnarray}
is obtained. We notice that the graphical procedure of solution requires 
operating in the three-dimensional space.
Analyzing the landscape $\xi({\bf p})$ we see that the interaction term 
does not affect the region close to the center $(0,0)$. At the same time, 
its involving gives rise to an elevation of the area adjacent to the 
van Hove points that eventually results in the appearance of the FC in 
this region since otherwise energy of the occupied states would exceed that 
of unoccupied ones. Results of numerical calculations are given in Fig.3,  
where the shadded areas correspond to the regions of the FC.

The models with limited long range forces are not exceptional. Similar
results are obtained, e.g., in a model with finite range forces given
by the Yukawa formula in the momentum space:
$H({\bf p}_1-{\bf p}_2)=g\exp (-\beta|{\bf p}_1-{\bf p}_2|)$
\cite{physrep,noz}.
Now the l.h.s.~of eq.~(\ref{eqnoz}) is recast as
\beq
L(\xi,p)\equiv (-\Delta+\beta^2)\,{\xi-\xi^0_p\over g}=n(\xi) .
\label{eqyuk}
\eeq
In this case, the graphical representation of solution is complicated
by the presence of the term $\Delta \xi$. Fortunately, in the FC region,
this term vanishes and we come back to eq.~(\ref{eqnoz}). Some
complifications emerge because of the existence of discontinuities in
the term $\Delta \xi$ at the boundaries of the FC region \cite{noz}
giving rise to jumps in the distribution $n(p)$ at these points.
The range of the jumps, which drops with increasing the coupling
constant $g$, at relatively small $g$ may attain 1 and then the FC
solution ceases to exist. As a result, requirements for the appearance
of the FC solution become more severe: the constant $g$ should exceed
a nonzero positive critical value $g_c$.

To make the next step, let us recast eq.~(\ref{main}) as follows
\beq
\int H^{-1}({\bf p},{\bf p}_1)\,\xi({\bf p}_1)\,d\tau_1-X_0({\bf p})=n(\xi),
\label{main1}
\eeq
where $X_0=(H^{-1}\xi^0_p)$. The model (\ref{eqyuk}) is recovered provided
$ H^{-1}({\bf p},{\bf p}_1)=g^{-1}(\beta^2-\Delta)\linebreak
\delta({\bf p}-{\bf p}_1)$. For a certain class of the momentum dependent
operators $H^{-1}$ this expansion can be extended to
\beq
H^{-1}({\bf p},{\bf p}_1)\,\xi({\bf p}_1)=(A_0-A_2\Delta+A_4\Delta^2\dots)
\,\delta({\bf p}-{\bf p}_1).
\label{expan}
\eeq
Upon substituting this equation into eq.~(\ref{main1}) one finds
\beq
(A_0-A_2\Delta+A_4\Delta^2\dots)(\xi-\xi^0_p)=n(\xi) .
\label{eqg1}
\eeq
Again all the terms containing the Laplace operators applied to $\xi$
vanish in the FC region where $\xi=0$, and the FC solutions ceases
to exist provided the jumps of $n$ at the boundaries of the FC region
attain the unity. Thus we infer that for a quite wide class of the
effective single particle Hamiltonian $H$, the FC solutions $\xi=0$
exists. The extension of this result to the general form of the
operator $H({\bf p},{\bf p}_1)$ is straightforward. We seek the FC
solutions in an equation
\beq
0=\xi^0_p+\int H({\bf p},{\bf p}_1;n)\,n(\xi({\bf p}_1))
\,d\tau_1, \qquad {\bf p}\in\Omega ,
\label{eqfc}
\eeq
obtained from eq.~(\ref{main}) setting there $\xi=0$.

It is the usual integral Fredholm equation of the first kind. As known,
numerical solving such equations encounters difficulties. Specialized
methods have been developed to overcome these difficulties which are
often called inverse problems (see e.g.~\cite {num}).
We are going to analyze them in a separate paper.

Thus we see that the FC solutions can be obtained proceeding from
the Landau-Pitaevskii relation between the single particle spectrum
$\xi({\bf p})$ and the quasiparticle momentum distribution $n({\bf p})$,
without addressing to the variational condition (\ref{var}).
What is remarkable is that the Migdal jump in the distribution function $n$
at the point $\xi=0$, being a building block of Fermi liquid
theory in systems with velocity independent forces, becomes a principal
factor for the appearance of  the FC solutions in systems where
such a dependence turns out to be sufficiently strong.

To conclude, having at hand the FC solution $\xi({\bf p}\in\Omega)=0$,
one can recover the variational condition (\ref{var}) suggesting
$\varepsilon({\bf p};n)\equiv \delta E_0(n)/\delta n({\bf p})$
and remembering that $\xi({\bf p})=\varepsilon({\bf p})-\mu$. In principle, 
the functional $E_0(n)$
can be found from this variational equation provided the
dependence of $\varepsilon({\bf p},n)$ on the distribution $n$
is supposed to be known.
\vskip 0.5 cm
\begin{centerline}
{\bf 3. The extension of the Landau formalism to marginal Fermi liquid}
\end{centerline}

\vskip 0.3 cm
If the systems with the FC  were subject to the Landau theory then
the development of the self-consistent theory of the fermion
condensation would be completed.
However, this is not the case. As we shall see, the mass operator
$\Sigma({\bf p},\varepsilon) $ in these systems contains the irregular
logarithmic term $\sim \varepsilon\ln|\varepsilon|$ inherent in marginal
Fermi liquid. This term makes it impossible  expand $\Sigma(\varepsilon)$ 
 in a Taylor series in energy. Although the pole in the
single particle Green function $G$ survives but the residue turns out
to be logarithmically small vanishing at the Fermi surface (see Fig.~4).

In spite of this fact, there is good reason to believe that
underlying ideas of the Landau theory remain robust including that of
the decomposition of the single particle Green function $G$
into a sum $G=G^r+ZG^s$ with a singular part $G^s$ (we shall call it
the pseudoparticle propagator) containing the logarithmic components
of the mass operator. In homogeneous matter, we write $G^s$ in a
simple form
\beq
G^s(p,\varepsilon)={Z^{-1}\over
\varepsilon-\sigma(\varepsilon)-s(p)-i\kappa(\varepsilon)} .
\label{gs}
\eeq
The connection between the quantities $\sigma$ and $\kappa$
stems from the dispersion relation (\ref{dis}) which
implies \cite{varma}
\beq
\kappa(\varepsilon\to 0)=
-\kappa\varepsilon,\qquad \sigma(\varepsilon\rightarrow 0)=
{2\over \pi}\kappa\varepsilon\ln{|\varepsilon|\over \varepsilon_L},
\label{marg}
\eeq
where $\kappa$ and $\varepsilon_L\sim \varepsilon^0_F=p^2_F/2M$
are numerical constants.

In the Landau theory, the quasiparticle distribution $n(\xi)$
is the universal function with the kink at $\xi=0$.
In marginal liquids, the pseudoparticle distribution function
$n(\xi)$ ceases to be universal. Analogously to eq.~(8), it
is evaluated from relation
$$
n(\xi)={1\over \pi}\int\limits_{-\infty}^{\infty}
{\rm Im} G^s({\bf p},\varepsilon)d\varepsilon
\qquad\qquad\qquad
$$
\beq
\equiv {1\over 2} +
{1\over 2\pi }\,\biggl[\,\int\limits_0^{\infty}{\gamma(\varepsilon)
 d\varepsilon\over [\varepsilon{-}\sigma_L(\varepsilon){-}
\xi]^2+\gamma^2(\varepsilon)}
-\int\limits_0^{\infty}{\gamma(\varepsilon)d\varepsilon\over
[\varepsilon{-}\sigma_L(\varepsilon){+}\xi]^2
+\gamma^2(\varepsilon)}\,\biggr],
\label{psp1}
\eeq
where we introduced notations $\xi=Zs,\gamma=Z\kappa$ and
$\sigma_L(\varepsilon)=
{2\over \pi}\gamma\varepsilon\ln{|\varepsilon|\over \varepsilon_L}$.
The constant $\varepsilon_L$ in (\ref{gs}) can be chosen in such a way
as to ensure the equality between the particle and the "pseudoparticle"
numbers, i.e.
\beq
\rho={\rm Tr}\int G^s({\bf p},\varepsilon)\,d\tau
{d\varepsilon\over 2\pi i}.
\label{psnorm}
\eeq
We notice that if the ratio $\gamma(\xi)/\xi$ is relatively small
then the integral (\ref{psp1}) is close to 1, provided $\xi<0$, or to 0
in case $ \xi>0$. On the other hand, if this ratio is of order of
unity (marginal regime) then there $n(\xi)$ essentially deviates
from $\theta(\xi)$.
As a result, the strength of the FC states turns out to be
distributed over all this interval and it is  its upper boundary
$\varepsilon_{FC}$ that should be compared to the characteristic FC
energy.

The renormalization of eq.~(\ref{id}) in the marginal region is
implemented along the same lines as before with the natural
redefinition of the singular part $A$ which now contains the product
$G^sG^s$. Furthermore in contrast to the quasiparticle approximation,
$\xi({\bf p})$ does not coincide with the spectrum of the single
particle excitations determined by the location of the pole of
$G^s({\bf p},\varepsilon)$ although both these quantities are related
to each other unambiguously. After some algebra one finds
\beq
{\partial \xi({\bf p})\over\partial p_{\alpha}}={p_{\alpha}\over M}
+\int {\cal F}({\bf p},{\bf p}_1) {\partial n({\bf p}_1)
\over \partial p_{1\alpha}}d\tau_1.
\label{lpm}
\eeq
In obtaining this result, we ignored the possible alteration of
the regular part $B$ due to corrections related to singular terms
in the mass operator. 

Again the integration of this equation yields
\beq
\xi({\bf p})={p^2\over 2M}+\int H({\bf p},{\bf p}_1)\,n({\bf p}_1)
\,d\tau_1.
\label{meq}
\eeq
However, in marginal Fermi liquid, the dependence of the distribution
$n(\xi)$ on $\xi$ is intricate and for finding the flat portion of the
spectrum, one needs to solve eqs.~(\ref{meq}) and (\ref{psp1})
self-consistently. The result crucially depends on the magnitude of
the marginal term in the mass operator $\Sigma$.

\newpage
\begin{centerline}
{\bf 4. The evaluation of the irregular part of $\Sigma$ beyond
the phase transition point}
\end{centerline}

\vskip 0.3 cm
The purpose of this chapter is to  calculate the imaginary part of
the mass operator $\Sigma({\bf p},\varepsilon)$ in the presence of
a flat portion in the single particle spectrum. In doing so,
the temperature $T$ is supposed to exceed a temperature $T_c$
of a phase transition, say, Cooper pairing or crystallization,
that results in lifting the degeneracy of $\xi({\bf p})$ implied by
the variational condition (\ref{var}). In this article, we investigate
behavior of the mass operator
$\Sigma({\bf p},\varepsilon)$ in the limit $T\ll\varepsilon$ and
therefore set $T= 0$. Without loss of generality, we restict
ourselves to the decay of a particle implying $\varepsilon>0$
in all the formulas. Different contributions to
${\rm Im}\,\Sigma({\bf p},\varepsilon>0)$ fall into two categories:
i) the particle decay into a particle and a particle-hole pair
and ii) the decay into a particle and a collective state.
The evaluation of the last contribution requires special calculations
of the collective spectra in the systems with the FC. This problem
is beyond the scope of this article. It will be studied later.
With these restrictions, ${\rm Im }\,\Sigma({\bf p},\varepsilon)$
is given by \cite {trio,pines}
$$
{\rm Im}\,\Sigma({\bf p},\varepsilon>0)=\int\!\!\!\int
d\tau d\tau_3\int\limits_0^{\varepsilon}{d\omega\over 2\pi}
\int\limits_0^{\omega}{d\varepsilon_3\over 2\pi}\,
W({\bf p},\varepsilon,{\bf p}_1,\varepsilon_1,
{\bf p}_2,\varepsilon_2, {\bf p}_3,\varepsilon_3)  $$
\beq
\times{\rm Im}\,G({\bf p}_1,\varepsilon_1)\,
{\rm Im}\,G({\bf p}_2,\varepsilon_2)\,
{\rm Im}\,G({\bf p}_3,\varepsilon_3).
\label{immas}
\eeq
The 4-momenta $p,p_2$ refer to incoming particles while $p_1,p_3$,
outgoing ones:
${\bf p}_3={\bf p}{+}{\bf p}_2{-}{\bf p}_1; \linebreak
\varepsilon_3=\varepsilon+\varepsilon_2-\varepsilon_1=\varepsilon_2+
\omega $. The limits of integration over $\varepsilon_3$ are imposed
by the restrictions: \linebreak
i) $\varepsilon_3>0$ since the 4-vector $p_3$
corresponds to the particle and ii) $\varepsilon_3<\omega$ since
$\varepsilon_2<0$ because the 4-vector $p_2$ corresponds to the hole.
In the three-dimensional system $d\tau =d^3q/(2\pi)^3$. But most of
experimental data on properties of strongly correlated Fermi systems
are obtained in studies of anisotropic quasi-two-dimensional crystals
where $d\tau=d^2 q/(2\pi)^2$. The transition probability $W$ is
related to the square of the effective interaction amplitude
between particles averaged over spin variables \cite{pines}
\beq
W={1\over 4}|\Gamma_{\uparrow\uparrow}|^2 +{1\over 2}
|\Gamma_{\uparrow\downarrow}|^2.
\eeq
Our prime goal is to establish the energy dependence of the mass
operator $\Sigma({\bf p}, \varepsilon)$ beyond the FC phase transition point
and further numerical factors will be omitted. In this article, we also 
neglect terms containing $G^r$ since their contributions
are reduced to a renormalization of input parameters.

Usually, the interaction $\Gamma$ varies slowly versus its arguments.
As a result, the integration is essentially facilitated \cite{pines}.
But it is not always true. We shall see that in the systems with the FC,
behavior of $\Gamma$ as a function of $\omega$ drastically changes that
has a dramatic impact on the energy dependence of the mass
operator $\Sigma$. At the same time, dealing with the damping effects
the momentum dependence of $\Gamma$ is of no crucial importance.
Therefore to avoid unjustified complifications we
restrict ourselves to the momentum-independent
zero Landau harmonic of the interaction amplitude.
Then $\Gamma_{\uparrow\uparrow}=0$ while $\Gamma_{\uparrow\downarrow}$
is expressed in terms of the scalar component $\Gamma$ of the
interaction amplitude $\Gamma_{\uparrow\downarrow}=2\Gamma$.
This suggestion being consistent with our choice (\ref{gs}) of
the form of $G^s$  furnishes the opportunity to
integrate over ${\bf p}_3$ and rewrite (\ref{immas}) in the form
\beq
\gamma({\bf p},\varepsilon)\sim\int\! d\tau\!
\int\limits_0^{\varepsilon} {d\omega\over 2\pi}
{\gamma({\bf p}{-}{\bf q},\varepsilon{-}\omega)\,
|\Gamma({\bf q},\omega)|^2\,
{\rm Im}\,A({\bf q},\omega)\over
[\varepsilon-\omega-\sigma_L({\bf p}{-}{\bf q},\varepsilon{-}\omega)-
\xi({\bf p}{-}{\bf q})]^2+
\gamma^2({\bf p}{-}{\bf q},\varepsilon{-}\omega)},
\label{kph}
\eeq
where ${\bf q}={\bf p}-{\bf p}_1$.

The interaction $\Gamma$ is related to the effective interaction potential
${\cal F}$ by the well known Landau equation
\beq
\Gamma({\bf q},\omega)={\cal F}({\bf q})+
{\cal F}({\bf q})\,A({\bf q},\omega)\,\Gamma({\bf q},\omega) =
{{\cal F}({\bf q})\over 1 -{\cal F}({\bf q})A({\bf q},\omega)},
\label{gamm}
\eeq
developed in the particle-hole (p-h) channel. As usual, the factor
$Z^2$ is already included into the definition of the effective
interaction $\Gamma$. Then the propagator
$A({\bf q},\omega)$ given by
\beq
A({\bf q},\omega)=2\int
G^s({\bf p},\varepsilon)\,G^s({\bf p}{-}{\bf q},
\varepsilon{-}\omega)\,d\tau\,{d\varepsilon\over 2\pi i}.
\eeq
is conveniently rewritten \cite{trio} as
$$
A({\bf q},\omega)\equiv N_n(0)\,
\Bigl(a({\bf q},\omega)+ib({\bf q},\omega)\Bigr),
$$
with $N_n(0)=p_FM^*/\pi^2$
being the density of states of the normal Fermi liquid and $M^*$,
the effective mass of the normal quasiparticle, and
\beeq
b({\bf q},\omega)&=&{1\over N_n(0)}
\int\!\!\int\limits_0^{\omega}
{\rm Im}\,G^s({\bf p},\varepsilon)\,
{\rm Im}\,G^s({\bf p}{-}{\bf q},\varepsilon{-}\omega)\,d\tau\,
{d\varepsilon\over \pi} ,
\nonumber\\
a({\bf q},\omega)&=&{1\over N_n(0)}\int\!\!\!
\int\limits_{-\infty}^{\infty}
{\rm Im}\,G^s({\bf p},\varepsilon)\,
{\rm Re}\,\Bigl[\,G^s({\bf p}{-}{\bf q},\varepsilon{-}\omega)+
G^s({\bf p}{-}{\bf q},\varepsilon{+}\omega)\,\bigr]
\,d\tau\,{d\varepsilon\over\pi}.
\label{prop}
\eeeq
In ordinary Fermi systems,
the magnitude of $|A|/N_n(0)$ is of order of 1.

In the strongly correlated Fermi systems, such as liquid He-3,
dense neutron matter or electrons in high-temperature
superconductors the amplitude ${\cal F}$ is repulsive
(${\cal F}>0$) and strong so that, as a rule, the term $|{\cal F}A|$
in the denominator of (\ref{gamm}) significantly exceeds the unity.
Omitting 1 in eq.~(\ref {gamm}), we are left with \cite{zeit}
\beq
\Gamma({\bf q},\omega)=-{1\over A({\bf q},\omega)} .
\label{str}
\eeq
In this limit,  the  interaction amplitude $\Gamma$
is completely
independent of  characteristics of the effective interaction
potential ${\cal F}$. It holds over a wide range of frequencies being
violated only close to lines $\omega_s(q)$ corresponding to
the collective excitations of the system.

Irregular contributions to $\Sigma$ may also emerge in the
particle-particle (p-p) channel, that is important, e.g.~in
superfluid systems. Here we study the systems for which the critical
temperature $T_c$ of the superfluid phase transition is very low
(e.g.~Sr$_2$RuO$_4$ with $T_c\simeq 1~ K$) and consider the case
$T>T_c$, at which the p-p correlations make no difference and
therefore below we omit these contributions.

In what follows we consider a crystal with a cubic or square lattice
assuming the FC to be positioned in the vicinity of
the van Hove points \cite {vol,clark}. We also set that the FC
density $\rho_c=\eta\rho$ is rather small: $\eta<1$.
The total damping $\gamma$ from (\ref{kph}) is decomposed into a sum
\beq
\gamma({\bf p},\varepsilon)=\gamma_0(\varepsilon)+
\gamma_1({\bf p},\varepsilon)+\gamma_2({\bf p},\varepsilon).
\label{sum}
\eeq
The term $\gamma_0$ does not contain the final FC Green functions.
Contributions containing the single final FC Green function
are included into the term $\gamma_1$, those with two FC ones,
into the last term $\gamma_2$. It should be indicated that for the
square or cubic lattice the decay of the FC particle into the final
state with no less than two FC states unambiguously results in
the appearance all three FC final states.

Dealing with the momentum dependence of $\gamma({\bf p})$, two regions
can be distinguished: i) the FC region $\Omega$ where the single particle
spectrum $\varepsilon({\bf p})$ is flat and ii) the ordinary region where
the gradient $d\xi/dp_n$ has the usual order of value $\simeq p_F/M^*_0$
where $M^*_0$ being independent of $T$ is the effective mass of the
ordinary quasiparticle. Further we shall neglect a possible variations of
the relevant quantities inside these regions and use the subscript $c$
for the FC region and the subscript $n$, for the "normal" one,
e.g.~$\gamma({\bf p}\in\Omega)\equiv\gamma_c$ or
$\gamma({\bf p}\notin\Omega)\equiv\gamma_n$.
We also use the notation $\Gamma_{00}$ for the set of diagrams of the
interaction amplitude without the FC states and the notation
$A_n({\bf q},\omega)$, for the particle-hole propagator of the normal
Fermi liquid.

In many respects, the evaluation of the quantities $\Gamma_{00}(q,\omega)$
and $A_n(q,\omega)$ at $q\sim p_F$ is equivalent to that in normal
homogeneous Fermi liquid. The momentum integration in Feynman
diagrams is carried out with the help of the standard formula
\beq
dp_xdp_ydp_z=dSdp_{\nu}={dSd\xi\over |d\xi/dp_{\nu}|},
\eeq
where $\nu$ indicates the normal to the Fermi surface direction.

Overwhelming contributions to $\gamma_0$ come from the momentum region
quite far from the van Hove points. Here the components $b_0$ and $a_0$
of the propagator $A_n(q,\omega)$ and the function
$|\Gamma_{00}(q,\omega)|^2$ are known to vary slowly enough that allows us 
to replace them by averaged values:
\beq
b_n\sim\omega{M^*\over p_F^2},\qquad
|\Gamma_{00}|^2\sim{1\over N_n^2(0)}\sim {1\over (M^*)^2}.
\label{aver}
\eeq
Upon substituting (\ref{aver}) into eq.~(\ref{kph}) the integrals
cease to depend on ${\bf p}$ at all, and we arrive at the common result
\beq
\gamma_0(\varepsilon)=-{\gamma_0 M^* \varepsilon^2\over p^2_F},
\label{kap0}
\eeq
where $\gamma_0$ is a dimensionless positive constant.

When evaluating $\gamma_1({\bf p},\varepsilon)$, two ordinary Green
functions entering eq.~(\ref{immas}) are combined into the particle-hole
propagator if the FC particle is generated, or into particle-particle
one when the FC particle is annihilated. As before, we
concentrate on the analysis of the contribution $\gamma_1^{ph}$.
Then the contribution $\gamma_{1n}$ is given by (see Fig.~5)
\beq
\gamma_{1n}(\varepsilon)\sim N_n(0)\int\limits_{\Omega}\!\!
\int\limits_0^{\varepsilon}d\tau_1 d\varepsilon_1
{\rm Im}\,G_c({\bf p}_1,\varepsilon_1)\,
|\Gamma_{01}({\bf p},{\bf p}_1,\varepsilon{-}\varepsilon_1)|^2\,
b_n(\varepsilon{-}\varepsilon_1),
\label{kph1}
\eeq
where
\beq
{\rm Im}G_c({\bf p}_1,\varepsilon_1)={\gamma_c({\bf p}_1,\varepsilon_1)
\over [\varepsilon_1-\sigma_c({\bf p}_1,\varepsilon_1)-
\xi_c({\bf p}_1)]^2+ \gamma^2_c({\bf p}_1,\varepsilon_1)}.
\eeq
The region $\Omega$ of the momentum integration in this formula consists
of "patches" of the FC adjacent to the van Hove points.

It can be verified that the interaction amplitude $\Gamma_{01}$ with
the single (final) FC state has the same order of value as
$\Gamma_{00}\sim 1/N_n(0)$. Then eq.~(\ref{kph1}) becomes
\beq
\gamma_{1n}(\varepsilon)\sim {M^*\over N_n(0)\,p^2_F}\,
\int\limits_{\Omega}\!\!\int\limits_0^{\varepsilon}
d\tau_1\,d\varepsilon_1\,
{\rm Im}\,G_c({\bf p}_1,\varepsilon_1)\,(\varepsilon_1{-}\varepsilon).
\label{kph11}
\eeq
In obtaining this result the relation (\ref{aver}) for the averaged value
$b_0 $ was used.

The integral (\ref{kph11}) considered versus energy $\varepsilon$,
the upper limit of integration, is, apparently, a monotonic function.
If the upper limit  were extended
to infinity, the integral value would be equal to $(1-n_0({\bf p}))$
provided $\xi({\bf p})>0$. In the opposite case $\xi({\bf p})<0$,
it is close to zero (recall, we consider the particle decay).
Integrating over momenta, one finds
\beq
\gamma_{1n}(\varepsilon)\sim {M^*\over N_n(0)\,p^2_F}
\int\limits_0^{\varepsilon}
N_c(\varepsilon_1)\,(\varepsilon_1{-}\varepsilon)\,d\varepsilon_1.
\eeq
The evaluation of this integral is facilitated if the energy
$\varepsilon$ exceeds the characteristic FC energy $\varepsilon_{FC}$.
In this case, in contrast to the density of the states $N_0(\varepsilon)$,
which varies slowly with $\varepsilon$, the density of the FC states
$N_c(\varepsilon)$ does extinct at $\varepsilon>\varepsilon_{FC}$.
Therefore it can be approximated by the $\delta$-function:
$N_c(\varepsilon)=\rho_c\delta(\varepsilon)$. As a result, at
$\varepsilon > \varepsilon_{FC}$ the FC contribution to $\gamma_n$
turns out to be linear function of energy rather than the quadratic
one as the ordinary term $\gamma_0$:
\beq
\gamma_{1n}(\varepsilon>\varepsilon_{FC})
   \simeq -\gamma_{1n}\,\eta\,\varepsilon, \qquad
\sigma_{1n} (\varepsilon>\varepsilon_{FC})\simeq
{2\gamma_{1n}\eta\over\pi}\,\varepsilon
\ln{|\varepsilon|\over\varepsilon_L},
\label{kap1}
\eeq
where $\gamma_{1n}$ is a constant of order unity.

As for the corresponding contribution  $\gamma_{1c}(\varepsilon)$
to the FC damping (see Fig.~6a), the equation for its evaluation
has the form
\beq
\gamma_{1c}(\varepsilon)\sim N_n(0)\int\limits_{\Omega}\!\!
\int\limits_0^{\varepsilon}d\tau_1\,d\varepsilon_1\,
{\rm Im}\,G_c({\bf p}_1,\varepsilon_1)\,
|\Gamma_{20}({\bf p},{\bf p}_1,
\varepsilon{-}\varepsilon_1)|^2\, b_n(\varepsilon{-}\varepsilon_1),
\label{kpc1}
\eeq
where $\Gamma_{20}$ stands for the amplitude of the transition of
the normal p-h pair into the FC pair.

The energy dependence of the quantities in this integral is more
important than the momentum one and therefore, in the first
approximation, the amplitude
$\Gamma_{20}({\bf p},{\bf p}_1,\varepsilon{-}\varepsilon_1)$ can be
replaced by the averaged over the FC region value
$\Gamma_{20}(\varepsilon{-}\varepsilon_1)$. Then after implementation
of the momentum integration one finds
\beq
\gamma_{1c}(\varepsilon)\sim N_n(0)
\int\limits_0^{\varepsilon} N_c(\varepsilon_1)|\,
     \Gamma_{20}(\varepsilon{-}\varepsilon_1)|^2\,
     b_n(\varepsilon{-}\varepsilon_1)\,d\varepsilon_1.
\label{kpc2}
\eeq

Two final FC states also come into play when the contribution
$\gamma_{2n}(\varepsilon)$ to the damping $\gamma(\varepsilon)$
is considered. Again we concentrate on the p-h contribution
(see Fig.~6b) given by
\beq
\gamma_{2n}(\varepsilon)
\sim N_n(0)\int \limits_{\Omega'}
\!\int\limits_0^{\varepsilon} d\tau_1\,d\varepsilon_1\,
  {\rm Im}\,G_n({\bf p}_1,\varepsilon_1)\,
  |\Gamma_{20}({\bf p},{\bf p}_1,\varepsilon{-}\varepsilon_1)|^2\,
  b_c({\bf p}{-}{\bf p}_1,\varepsilon{-}\varepsilon_1) .
\label{knc1}
\eeq
We introduced a special notation $\Omega'$ for the integration region
in (\ref{knc1}), determined by the requirement for the difference
${\bf p}-{\bf p}_1$  to have the value appropriate for the pair of
the FC states.
Neglecting the momentum dependence of the amplitude $\Gamma_{20}$
and the propagator $b_2$
this equation is recast to
\beq
\gamma_{2n}(\varepsilon)
\sim N_n^2(0) \int\limits_0^{\varepsilon}\,
|\Gamma_{20}(\varepsilon{-}\varepsilon_1)|^2\,
b_c(\varepsilon{-}\varepsilon_1)\,d\varepsilon_1.
\label{knc2}
\eeq

At last, dealing with the contribution $\gamma_{2c}(\varepsilon)$,
all four states belong to the FC (see Fig.~6c), so that
\beq
\gamma_{2c}(\varepsilon)\sim N_n(0)\int\limits_{\Omega}\!\!
\int\limits_0^{\varepsilon} d\tau_1\,d\varepsilon_1\,
{\rm Im}\,G_c({\bf p}_1,\varepsilon_1)\,
|\Gamma_{22}({\bf p},{\bf p}_1,\varepsilon{-}\varepsilon_1)|^2\,
b_c({\bf p}{-}{\bf p}_1,\varepsilon{-}\varepsilon_1).
\label{kcc1}
\eeq
This integral is reduced to
\beq
\gamma_{2c}(\varepsilon)\sim N_n(0)
\int\limits_0^{\varepsilon} N_c(\varepsilon_1)\,
|\Gamma_{22}(\varepsilon{-}\varepsilon_1)|^2\,
b_c(\varepsilon{-}\varepsilon_1)d\varepsilon_1.
\label{kcc2}
\eeq

The averaged over the FC region propagator $b_c(\omega)$ entering
this formula is defined in terms of
$$
b_c({\bf q},\omega)\sim {1\over N_n(0)}\int\limits_\Omega\!
\int\limits_0^{\omega} {\rm Im}\,G^s({\bf p_1},\varepsilon)
{\rm Im}\,G^s({\bf p}_1{-}{\bf q},\varepsilon{-}\omega)\,
d\tau\,d\varepsilon \qquad\qquad
$$
\beq
{}\sim {1\over N_n(0)}\int\limits_\Omega\!\int\limits_0^{\omega}d\tau
{\gamma_c(\varepsilon)\,\gamma_c(\varepsilon{-}\omega)\,d\varepsilon
     \over
\Bigl([\varepsilon{-}\sigma_c(\varepsilon){-}\xi_c({\bf p}_1)]^2+
\gamma^2_c(\varepsilon)\Bigr)
\Bigl([\varepsilon{-}\omega{-}\sigma_c
         (\varepsilon{-}\omega)-\xi_c({\bf p}_1{-}{\bf q})]^2+
 \gamma^2_c(\varepsilon{-}\omega)\Bigr)}.
\label{b22}
\eeq

Primarily, we study the case $\varepsilon\geq \varepsilon_{FC}$.
In this situation, the integrand in eq.~(\ref{b22}) is made up of
two separate peaks. The first is positioned at
$\varepsilon \sim \varepsilon_{FC}$ and contributes only if
$\xi_c({\bf p}_1)>0$ (i.e.~if $n_0({\bf p}_1)<1/2$) independently
of the sign of the other term $\xi_c({\bf p}_1)$. The second is
located at $(\omega-\varepsilon)\sim \varepsilon_{FC}$ and
contributes provided  $\xi_c({\bf p}{-}{\bf q})<0$
(i.e.~$n_0({\bf p}{-}{\bf q})>1/2)$ independently of $\xi_c({\bf p}_1)$.
Both these terms supplement each other and we arrive at
$$
b_c(\omega>\varepsilon_{FC})\sim{\eta\rho\over N_n(0)}\,
{\gamma_c(\omega)\over \Bigl([\omega-\sigma_c(\omega)]^2 +
\gamma^2_c(\omega)\Bigr)}
\biggl[\,\int\limits_0^{\omega}
{\gamma_c(\varepsilon{-}\omega)\,d\varepsilon \over
[(\varepsilon-\omega-\sigma_c(\varepsilon{-}\omega)]^2+
\gamma^2_c(\varepsilon{-}\omega)}
$$
\beq
+ \int\limits_0^{\omega}
{\gamma_c(\varepsilon)\,d\varepsilon \over
[(\varepsilon-\sigma_c(\varepsilon)]^2+
\gamma^2_c(\varepsilon)}\,\biggr].
\label{b21}
\eeq
Again we can use the fact that the integrands on the r.h.s.~of this
equation are $\delta$-like functions. Then after simple integration
we are left with
\beq
b_c(\omega)\sim\eta\,{p^2_F\over M^*}\,
{\gamma_c(\omega)\over f^2(\omega)},
\label{b20}
\eeq
where
\beq
f^2(\omega)=[\omega-\sigma_c(\omega)]^2 +\gamma^2_c(\omega).
\label{def}
\eeq
To correctly estimate the damping one also needs to know the behavior
of the real part $a_c$ of the propagator $A_c$ averaged over the FC
region. It can be traced from the formula
$$
a_c({\bf q},\omega)
\sim {1\over N_n(0)}\int\limits_\Omega\!
\int\limits_{-\infty}^{\infty}
{\rm Im}\,G^s({\bf p}_1,\varepsilon)\,
{\rm Re}\,\Bigl[\,G^s({\bf p}_1{-}{\bf q},\varepsilon{-}\omega)+
G^s({\bf p}_1{+}{\bf q},\varepsilon{+}\omega)\,\Bigr]\,
d\tau\,d\varepsilon
$$
$$
\sim {1\over N_n(0)}\int\limits_\Omega\!
\int\limits_{-\infty}^{\infty}d\tau\,
{\gamma_c(\varepsilon)\,d\varepsilon \over
[\varepsilon-\sigma_c(\varepsilon)-\xi_c({\bf p}_1)]^2
+\gamma^2_c(\varepsilon)} \qquad\qquad\qquad\qquad
$$
\beq
{\times}\biggl[{\varepsilon{-}\omega{-}\sigma_c(\varepsilon{-}\omega)
\over
[\varepsilon{-}\omega{-}\sigma_c(\varepsilon{-}\omega){-}\xi_c
({\bf p}_1{-}{\bf q})]^2{+}\gamma^2_c(\varepsilon{-}\omega)}
+
{\varepsilon{+}\omega{-}\sigma_c(\varepsilon{+}\omega)\over
[\varepsilon{+}\omega{-}\sigma_c(\varepsilon{+}\omega){-}\xi_c
({\bf p}_1{+}{\bf q})]^2{+}\gamma^2_c(\varepsilon{+}\omega)}\biggr].
\label{a22}
\eeq
After simple algebra, the following estimate for the averaged value
$a_c(\omega)$ is obtained
\beq
|a_c(\omega)|\sim {\eta\, p^2_F\,\varepsilon_{FC}\over
M^*\,\Bigl([\omega-\sigma_c(\omega)]^2 +
\gamma^2_c(\omega)\Bigr)}.
\eeq
Comparing $|a_c|$ to $|b_c|$ and bearing in mind that the value of
$\varepsilon_{FC}$ depends by itself on $\gamma_c$ we infer that, in
the first approximation, the $a_c$ contribution can be neglected.

It is worth noting that the integrands of (\ref{kpc2}) and (\ref{kcc2})
are peaked at $\varepsilon_1\sim \varepsilon_{FC}$ due to the
$\delta$-like behavior of ${\rm Im}\,G_c({\bf p}_1,\varepsilon_1)$ and
rather smooth behavior of the products $ b_n|\Gamma_{20}|^2$ and
$b_c|\Gamma_{22}|^2$. To verify this assertion consider the set
of equations for the amplitudes $\Gamma_{22}$, $\Gamma_{02}$.
In the symbolic form, it reads as
\begin{eqnarray}
\Gamma_{22}&=&{\cal F}_{22}+{\cal F}_{22}A_c \Gamma_{22}+
{\cal F}_{20}A_n\Gamma_{02} \nonumber\\
\Gamma_{02}&=&{\cal F}_{02}+{\cal F}_{02}A_c\Gamma_{22}+
{\cal F}_{00}A_n\Gamma_{02}.
\end{eqnarray}
This system is easily solved
\begin{eqnarray}
\Gamma_{20} &=&{{\cal F}_{20}\over
(1{-}{\cal F}_{00}A_n)(1{-}{\cal F}_{22}A_c)-{\cal F}_{20}^2A_cA_n},
\nonumber\\
\Gamma_{22}&=&  {{\cal F}_{22}(1{-}{\cal F}_{00}A_n)+{\cal F}_{20}^2A_n
\over 1-{\cal F}_{00}A_n- A_c[{\cal F}_{22}(1{-}{\cal F}_{00}A_n)+
{\cal F}_{20}^2A_n]}.
\label{sup}
\end{eqnarray}

 From the last formula we can infer that in the strong coupling limit
the value $|A_c\Gamma_{22}|$ is of order of 1.
In what follows we employ an interpolation formula
\beq
N_n^2(0)\,|\Gamma_{22}(\varepsilon)|^2
  \sim {1\over 1+ b^2_c(\varepsilon)}.
\label{gm}
\eeq
which is correct for a moderate interaction and in the strong coupling
limit, as well.  As for the ratio
$|\Gamma_{20}/\Gamma_{22}|\sim {\cal F}_{20}/{\cal F}_{22}$, its value
is, probably, considerably less than 1. Upon substituting these results
into eqs.~(\ref{kpc2}) and (\ref{kcc2}) we are led to the
following equations
\beq
\gamma_{1c}(\varepsilon)\sim -\eta\,
            {\varepsilon\over 1{+}b_c^2(\varepsilon)},  \qquad
\gamma_{2c}(\varepsilon)\sim \eta\,p^2_F\,
     {b_c(\varepsilon)\over M^*\,(1{+}b_c^2(\varepsilon))}.
\label{kpcr}
\eeq
As a result, the equation for finding $\gamma_c(\varepsilon)$ takes
the form
\beq
\gamma_c(\varepsilon)= - \gamma_0
{M^*\varepsilon^2\over p^2_F}-
 C\,\eta\,{\varepsilon\over 1{+}b_c^2(\varepsilon)}
+ A\,\eta\,p^2_F\,{b_c(\varepsilon)\over M^*\,(1{+}b_c^2(\varepsilon))}.
\label{eqsum}
\eeq

Upon substituting here the formula (\ref{b20}) for the propagator
$ b_c$ and simple transformations one obtains
\beq
\gamma_c(\varepsilon)\biggl(1-A{\delta^2_F f^2(\varepsilon)\over
f^4(\varepsilon)
+B^2\delta^2_F\gamma^2_c(\varepsilon)}\biggr)=
-\gamma_0{\varepsilon^2\over
\varepsilon^0_F}-C\eta{\varepsilon f^4(\varepsilon)\over
f^4(\varepsilon)+B^2\delta_F^2\gamma^2_c(\varepsilon)},
\label{intr}
\eeq
where $\delta_F=\eta p^2_F/M^*$ while the constants $A, B,C$ are of
order of 1. At $\varepsilon>\delta_F$, the magnitude of the propagator
$b_c$ is less than 1 and therefore the second term in the brackets
of eq.~(\ref{intr}) is relatively small. This implies
\beq
\gamma_c(\varepsilon>\delta_F)\sim -C\eta\varepsilon,
\label{damp1}
\eeq
where the constant $C$ is of order of 1. We infer that at these
energies the magnitude of the marginal terms in the mass operator
is quite small. It should be indicated that, according to (\ref{psp1}),
the presence of the marginal term in $\Sigma(\varepsilon\to 0)$,
no matter how small it is, alters the Landau distribution $n(\xi)$
giving rise to the an inclination of the plateau in the single
particle spectrum proportional to $\eta$ if this behavior persists
in the limit $\varepsilon\to 0$.

However, this smallness cannot always hold as $\varepsilon$ together
with $f(\varepsilon)$ go down to zero, otherwise the sum of the terms
in the brackets on the l.h.s.~of eq.~(\ref{intr}) would change its
sign. This circumstance makes the function $\gamma(\varepsilon)$
rapidly enhance so as to both the terms in the brackets almost cancel
each other approaching $\varepsilon\simeq \delta_F$. Indeed, let us
seek the solution of a simplified equation
$$
1-A\,{\delta^2_F f^2(\varepsilon)\over f^4(\varepsilon)
+B^2\delta^2_F\gamma^2_c(\varepsilon)}=0.
$$
It can be easily verified that no solution of this equation exists
at $\varepsilon$ exceeding a critical value
$\varepsilon_{cr}=\delta_F\sqrt{A}$ and $\gamma_c(\varepsilon_{cr})=0$.
As $\varepsilon$ goes down, $\gamma_c(\varepsilon)$ grows up so that
the ratio $r_c=\gamma_c(\varepsilon)/\varepsilon$ attains in the
region $\varepsilon\sim \delta_F$ values of order of the unity in
contrast to perturbation theory estimate $r_c\simeq \eta\ll 1$.
The alteration of behavior of the damping from the perturbative
regime (\ref{damp1}) to a "strong coupling limit"
\beq
\gamma_c(\varepsilon)\simeq\varepsilon, \qquad \varepsilon<\delta_F,
\label{damp2}
\eeq
occurs in a quite narrow region of energy that is confirmed in
numerical calculations.

The result $r_c(\delta_F)\sim 1$ is crucial for finding out behavior of
$\gamma_c(\varepsilon\to 0)$ where  a scaling
approach can be employed. We assume that
$\gamma_c(\varepsilon\to 0)\sim \varepsilon^{\nu}$. For 
the evaluation of  the critical index $\nu$, we calculate the propagator $b_c$
proceeding from an assumption that the damping term exceeds all
the rest ones in ${\rm Im}\,G_c({\bf p},\varepsilon)$. Then
$$
b_c(\omega)\sim\eta{\omega\over \gamma^2_c(\omega)}.
$$
This function diverges at $\omega\to 0$ provided $\nu>1/2$ while
the product
$$
|\Gamma_{22}(\varepsilon)|^2\,b_c(\varepsilon)\sim
{\gamma_c^2(\omega)\over \eta\,\omega}.
$$
Then the folding with ${\rm Im}\,G_c({\bf p},\varepsilon)\sim
1/\gamma_c$ leads to the strong-coupling-limit equation for
$\gamma_c$ which reads as
\beq
\gamma_c(\varepsilon\to 0)\simeq\int\limits_0^{\varepsilon}
\gamma_c(\omega)\,{d\omega\over\omega}.
\label{scal}
\eeq
Significantly that $\eta$ does not enter this equation. This allows us
to find the correct solution $\gamma(\varepsilon)\simeq \varepsilon$
which simultaneously meets eq.~(\ref{scal}) and the matching condition
$$
r_c(\delta_F)\simeq 1.
$$
Thus the marginal behavior of the mass operator $\Sigma(\varepsilon)$
in the system with the FC holds over a quite wide region of energy
up to $\varepsilon\sim \delta_F$. This means that beyond the FC phase
transition point there is no room for the Landau theory in the
vicinity of the origin $\varepsilon=0$. Otherwise the ratio
$r_c(\varepsilon\to 0)$ should vanish and then some completely flat
portion $\xi=0$ in the spectrum of the single particle excitations
should survive. But in this case calculations along the same lines as
done above results in the nonzero value of $r_c(\varepsilon\to 0)$.
This contradiction proves the inapplicability of the Landau theory.

Now we return to the damping $\gamma_n$. From eq.~(\ref{knc2}) one has
\beq
\gamma_{2n}(\varepsilon)=C_{2n} \int\limits_0^{\varepsilon}
 {b_c(\varepsilon_1)\over 1+b^2_c(\varepsilon_1)}\,d\varepsilon_1
\eeq
with the constant $C_{2n}$ whose value is less than 1. In principle,
this integral is a linear in $\varepsilon$ function in the region
$\varepsilon\simeq \delta_F$ so that the total damping
$\gamma_n(\varepsilon)$ is
\beq
\gamma_n(\varepsilon)=-\gamma_n\varepsilon,
        \qquad \varepsilon\leq \delta_F,
\label{dampn}
\eeq
the value of the constant $\gamma_n$, proportional to
${\cal F}_{20}/{\cal F}_{22}$ being, probably, considerably less than 1.

To establish the structure of the spectrum of the single particle
excitations beyond the FC phase transition point one needs to solve
self-consistently the system (\ref{psp1}) and (\ref{meq}) accounting
for the energy dependence of the damping $\gamma(\varepsilon)$ given
by eqs.~(\ref{damp1}), (\ref{damp2}) and (\ref{dampn}). One of the
ways to obtain the solution is the iterative procedure. We start
with the FC distribution $n({\bf p}) $ found from eq.~(\ref{eqfc})
assuming the explicit form of this equation, where the damping is
completely ignored, to be known. With this distribution at hand,
we evaluate the parameters determining the damping at different
energies. Having found these constants we substitute them into
eq.~(\ref{psp1}), evaluate the new distribution function $n({\bf p})$
and insert it into eq.~(\ref{meq}) to find the new single particle
spectrum and then repeat all the procedure till its convergence.

It should be kept in mind that the results depend on $\eta$ as an
input parameter. What actually happens is that the FC density
$\rho_c$ is not the input parameter. It is given by an integral
\beq
\rho_c=\int\limits_\Omega n({\bf p})\,d\tau.
\label{rhoc}
\eeq
The integration in (\ref{rhoc}) is performed over regions where
the slope of the single particle spectrum is small compared to
the usual one. In the quasiparticle approximation, the boundaries
of this region are determined unambiguously, since this slope is
identically 0, and the density $\rho_c$ is also evaluated
unambiguously being proportional to the excess
${\cal F}-{\cal F}_{cr}$. However, the damping implicitly influences
on the size of the FC region so that the integral (\ref{rhoc}) turns
out to be an equation for obtaining $\rho_c$ and with it the set
of equations describing the basic properties of the system beyond
the FC phase transition point becomes complete.

The numerical solution of this system is beyond of the scope of this
article. We restrict ourselves to several comments.
Since the ratio $\gamma_c(\varepsilon)/\varepsilon$ exceeds 1 from
0 to $\delta_F$ the kink at $\xi=0$ in the distribution $n(\xi)$
evaluated in the quasiparticle approximation is split: its upper part
moves to the left while the low one, to the right and the magnitude of
this shift is of order of $\delta_F$. Steep pieces of the curve
$n(\xi)$ are connected by a bridge the length of which and the slope
depend on input parameters. If it is almost horizontal, then the
insertion of such a distribution $n(\xi)$ into eq.~(\ref{eqfc})
results in splitting the plateau in the single particle spectrum,
one part of which is shifted upward while the other, downward of the
Fermi level resembling a zigzag. The lower part occupied by electrons
can be investigated in ARPES studies while the upper one, not.
In this case, the spectrum of the former FC states remains practically
flat but it is shifted from the Fermi surface. Its value
$\simeq \delta_F$ can be treated as the characteristic FC energy
$\varepsilon_{FC}$.

On the other hand, if the slope of the bridge is not so small
then the zigzag gets smeared. Then the average slope of the spectrum
is estimated from an approximate formula
$$
\varepsilon_{FC}\simeq \int\limits_\Omega {d\xi\over dp}\,dp
\simeq {\overline {d\xi\over dp}}\,L.
$$
In the two-dimensional problem, the characteristic size $L$ of the
FC region is proportional to $\eta^{1\over 2}$, in the
three-dimensional system it is proportional to $\eta^{1\over 3}$.
Substitution here $\varepsilon_{FC}\simeq\delta_F$ yields in the
two-dimensional case the result $d\xi/dp\sim \eta^{1\over 2}$ while
in the three-dimensional system $d\xi/dp\sim \eta^{2\over 3}$.
Thus we infer that the plateau in the single particle spectrum is
inclined but the slope turns out to be small:
it vanishes together with the condensate density $\rho_c$.

\vskip 0.5 cm
\begin{centerline}
{\bf 5. Comparison with the phenomenological theory by
Varma et al.~\cite{varma}}
\end{centerline}

\vskip 0.3 cm
Now we are able to compare the results obtained with postulates
of the phenomenological theory by Varma et al.~\cite {varma} of normal
states of high-$T_c$ superconductors. The first postulate is concerned
with the polarizability $P({\bf q},\omega)$ and requires it to be an
$\omega$-independent quantity in contrast to the ordinary linear
$\omega$-dependence. The second is related to the mass operator
$\Sigma({\bf p},\varepsilon)$. It is assumed to have the form
(\ref{marg}) inherent in marginal Fermi liquid. As we have seen,
such a behavior of the mass operator $\Sigma$ beyond the FC phase
transition point is due to flattening the single particle spectra.

As for the polarizability $P({\bf q},\omega)$ which determines
a relation between an external field $V_0$ and a real one $V$ acting
inside of the system:
$V({\bf q},\omega)=P({\bf q},\omega)\,V_0({\bf q},\omega)$, we
evaluate it neglecting  connections between the normal and the FC
regions, i.e.~suggesting $\Gamma_{20}\equiv 0$. Then \cite {mig}
$P^{-1} ({\bf q},\omega)=1-{\cal F}({\bf q})A_n({\bf q},\omega)$ and

\beq
{\rm Im}\,P(q,\omega)={\rm Im}\,A_n(q,\omega)\,|P(q,\omega)|^2
\simeq N_n(0)\,
{b_n(q,\omega)\over |1-{\cal F}(q)A_n( q,\omega)|^2}.
\label{pol}
\eeq
In the optical conductivity problem, the value of the wave vector
$q=\omega/c$ is so small that the damping effects exhibit themselves
in full force. In the normal region,
\beq
b_n(q{=}0,\omega)\sim {1\over N_n(0)}
\int\!\!\!\int\limits_0^{\omega}
{\gamma_n(\varepsilon)\gamma_n(\varepsilon{-}\omega)\,d\varepsilon
\,dp_{\nu}\,dS \over
\Bigl([\varepsilon{-}\sigma_n(\varepsilon){-}\xi_n({\bf p})]^2+
\gamma^2_n(\varepsilon)\Bigr)
\Bigl([\varepsilon{-}\omega{-}\sigma_n
    (\varepsilon{-}\omega){-}\xi_n({\bf p})]^2+
\gamma^2_n(\varepsilon{-}\omega)\Bigr)}.
\label{im0}
\eeq
This integral is evaluated analytically and the result
\beq
b_n(q\to 0,\omega) \simeq - \int\limits_0^{\omega}
{d\varepsilon \over
|\gamma_n(\varepsilon)|+|\gamma_n(\varepsilon{-}\omega)|} .
\label{imb0}
\eeq
is drastically enhanced compared to the conventional one (\ref{aver}),
proportional to $\omega$. Upon substituting into (\ref{im0}) the
result $\gamma_n(\varepsilon)=-\gamma_n\varepsilon$ and simple
integration one finds
\beq
b_n(q\to 0,\omega)\simeq - {1\over \gamma_n}
\label{ib0}
\eeq
where $\gamma_n\sim 1$. With this result, simple algebra leads to the
following estimate
\beq
{\rm Im}\,P(q<q_{cr},\omega)\sim -N_0(0).
\label{ipol}
\eeq
Thus the first postulate of the theory \cite {varma} consisting of
the independence of the polarizability $P(\omega)$ of the frequency
$\omega$ is reproduced in our analysis, too.

With increasing $\omega$ the damping grows up and the magnitude of
${\rm Im}\,P$ drops. As $\omega$ increases from $\varepsilon_{FC}$
the damping $\gamma(\varepsilon)$ rapidly drops. As a result,
the value of $b_n(q\to 0,\omega)$ grows up. However, simultaneously
the denominator of (\ref{pol}) increases, even faster than the
numerator, and
${\rm Im}\,P(q\to 0,\omega)$ falls rapidly to values of order of
$\eta N_0(0)$. As a result, the optical conductivity of the
system with the FC has the maximum at frequencies
$\omega\simeq \eta\varepsilon^0_F$. Thus in the systems with the FC
there exists a new mechanism, different from the polaron one,
for the appearance of irregular behavior of the optical
conductivity $\sigma(\omega)$.

The behavior of the real part $a_n( q,\omega)$ in the static limit
also differs from the Fermi liquid one. Indeed
\beq
a_n(q{=}0,\omega{=}0) \sim {1\over N_n(0)}
\int\!\!\!\int\limits_{-\infty}^{\infty}
{\gamma_n(\varepsilon)\,
[\varepsilon{-}\sigma_n(\varepsilon){-}\xi_n( p)]\,
d\varepsilon\,d^3p \over
\Bigl([\varepsilon{-}\sigma_n(\varepsilon){-}\xi_n( p)]^2+
\gamma^2_n(\varepsilon)\Bigr)^2}.
\label{ima0}
\eeq
In contrast to the quasiparticle-approximation formula, this integrand has 
no singularities and therefore 
replacing integration over $p$ by that over
$t=\varepsilon{-}\sigma_n(\varepsilon){-}\xi_n( p)$ and assuming the
derivative $dt/dp=-d\xi_n(p)/dp$ to be constant one finds
\beq
{\rm Re}\,A(q{=}0,\omega{=}0)=0.
\eeq
The Landau result is restored only if $q>q_{cr}$. Of course, this result 
holds only if the momentum dependence of the group velocity $\xi_n(p)$ is 
negligible but usually this dependence is sufficiently weak that results 
in a significant suppression of the real part of the particle-hole propagator 
and, in its turn, in the suppression of the well known results for the spin 
and charge susceptibility. From these results one obtains that the well 
known singular behavior of the propagator $A(q,\omega)$ in the static limit 
$q\to 0,\omega\to 0$ is absent in the systems with the FC and the reason for 
that is the linear in $\varepsilon$ damping of the single particle degrees of 
freedom.

\vskip 0.5 cm
\begin{centerline}
{\bf 6. Conclusion}
\end{centerline}

\vskip 0.3 cm
In this article, proceeding from the standard many-body-theory
formalism we analyzed the interplay between the flattening and
the damping in the single particle spectra of the strongly correlated
Fermi systems beyond the FC phase transition point assuming the FC
density $\rho_c$ to be relatively small: $\eta=\rho_c/\rho<1$.
We concentrated on crude features of this phenomenon not caring about
numerical factors in quantities of our interest. We have found that
in the quasiparticle approximation there exists a class of solutions
of the many body theory equation for the spectrum $\xi(p)$ with the zero 
value of the quasiparticle
group velocity at the Fermi surface. The necessary condition for the
appearance of these solutions is a sufficiently strong momentum
dependence of the mass operator $\Sigma({\bf p},\varepsilon)$.
Such a strong momentum dependence may stem from the fact that in the
vicinity of many phase transition points, e.g.~antiferromagnetism
the effective interaction $\Gamma({\bf q})$ has quite narrow peaks. 
Employing the RPA for the evaluation of the contribution of these resonance 
components into the mass operator $\Sigma(p,\varepsilon)$ a well pronounced 
momentum dependence does emerge in this function. Thus we can infer that
in strongly
correlated electron systems of solids, the fermion condensation could
be entailed by the spin fluctuations and, hence, the fermion
condensation is the precursor of antiferromagnetism.

We demonstrated that in the vicinity of the Fermi surface there is no
room for the Landau theory: the system with the FC does behave itself
as marginal Fermi liquid, i.e.~the damping $\gamma(\varepsilon)$ of
the states adjacent to the Fermi surface is linear in energy.     
The enhancement of $\gamma$ should entail a significant spread
of the FC states and influences on the intensity of the peaks
observed in ARPES studies in the region occupied by the FC. This
intensity is proportional to the factor
$\Bigl(1{-}{\partial\Sigma({\bf p},\varepsilon)\over\partial
\varepsilon}|_{\varepsilon=\varepsilon({\bf p})}\Bigr)^{-1}$ and
therefore it is suppressed in the FC region adjacent to the van Hove
points. Of course, the magnitude of this suppression greatly depends
on input parameters and varies from one compound to another.

Two different marginal regimes of the energy dependence of $\gamma$
have been established. The perturbative regime
$\gamma(\varepsilon)\sim\eta\varepsilon$ is adequate at energies
$\varepsilon > \delta_F\simeq\eta\varepsilon^0_F$. The strong coupling
regime with $\gamma(\varepsilon)\simeq \varepsilon$ is involved in the
vicinity of $\varepsilon=0$. The transition from the first to the
second one is, probably, quite smooth, but dealing with the states
near the van Hove points this transition could be well pronounced
providing an unaccustomed shape of the energy dependence of
$\gamma(\varepsilon)$.

We established that behavior of the imaginary part of the polarization
operator \linebreak $P(q=0,\omega)$ of the system with the FC drastically
differs from that assigned by the Fermi liquid theory: its magnitude
remains constant over the low-frequency region. This effect can lead to
anomalies in the optical conductivity $\sigma(\omega)$ conventionally
associated with the polaron mechanism. We also uncovered an alteration 
of the real part of the particle-hole propagator that entails a suppression 
of the spin and charge susceptibility of the strongly correlated systems.

It is worth noting that the presence of the marginal term in the mass 
operator should change many formulas of the BCS theory of superfluid 
systems with the FC including the BCS relation between $\Delta(T=0)$ and 
the critical temperature $T_c$. 
This problem will be investigated in the next article in more detail.

In conclusion, that the analysis of many experimental data on the base of
Luttinger liquid where the renormalization constant $z=0$ is an appropriate 
guide to understanding of behavior of stronglycorrelated electron systems 
became clear, probably, ten years ago \cite {varma,and}. At the same
time, many properties of these systems are known to be succesfully 
treated proceeding from Fermi liquid theory. What we demonstated in our 
article is that the fermion condensation model serves as a "bridge" 
connecting both these approaches.    

We acknowledge M.~R.~Norman, H.~Ding and  G.~Kotliar  for 
valuable discussions of the above topics. We are also grateful to
A.~M.~Bratkovsky, D.~Dessau and  N.~E.~Zein.
V.~A.~Khodel thanks Prof.~J.~W.~Clark for the kind hospitality at
Washington University in St.~Louis. This research was supported by NSF Grant 
PHY-9900713 and by Mc Donnell Center for the Space Sciences (VAK).

\newpage
\centerline{\bf Figures captions}

\vskip 0.8 cm
Fig.~1. Graphical solution of eq.~(\ref{eqnoz}). The distribution
$n(\xi)$ (bold line) is crossed by the set of straight lines
$(\xi-\xi_p^0)/V$. The crossing points of these lines with the vertical line
 $\xi=0$ are included into the FC region $\Omega$.

%
\vskip 0.2 cm
Fig.~2. The graphical representation of the momentum dependent 
contribution to the mass operator (\ref{sigex}) resulting from 
exchange spin fluctuation term (\ref{exch}) of the effective 
interaction.

\vskip 0.2 cm
Fig.~3. FC regions in the square lattice within the model with
the spin-spin effective interaction (\ref{exch}) peaked at the 
vector ${\bf Q}=(\pi,\pi)$. The formfactors $\phi(\bf p)$
are assumed to vanish far from the van Hove points. 
The parameters used are: $t{=}0.7$, $\mu{=}1.1\varepsilon_0 t$, 
$\lambda{=}0.4\varepsilon_0$.
%

\vskip 0.2 cm
Fig.~4. The residue of the single-particle Green function in
marginal Fermi liquid 
with $\kappa(\varepsilon)=-\varepsilon$
as a function of energy $\varepsilon$ measured in 
$\varepsilon_F^0$.

\vskip 0.2 cm
Fig.~5. The graphical representation for the contribution $\gamma_{1n}$ 
to the total damping.

\vskip 0.2 cm
Fig.~6. The same for $\gamma_{1c}$ (a), $\gamma_{2n}$ (b)
and $\gamma_{2c}$ (c).

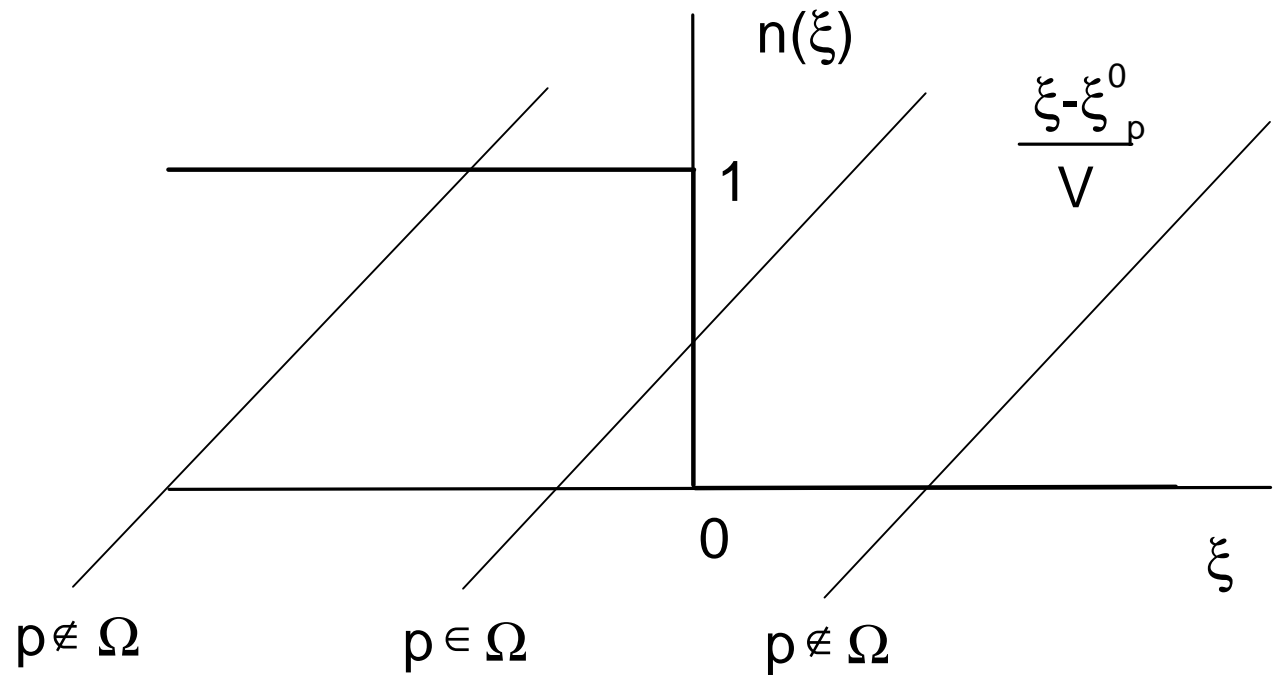
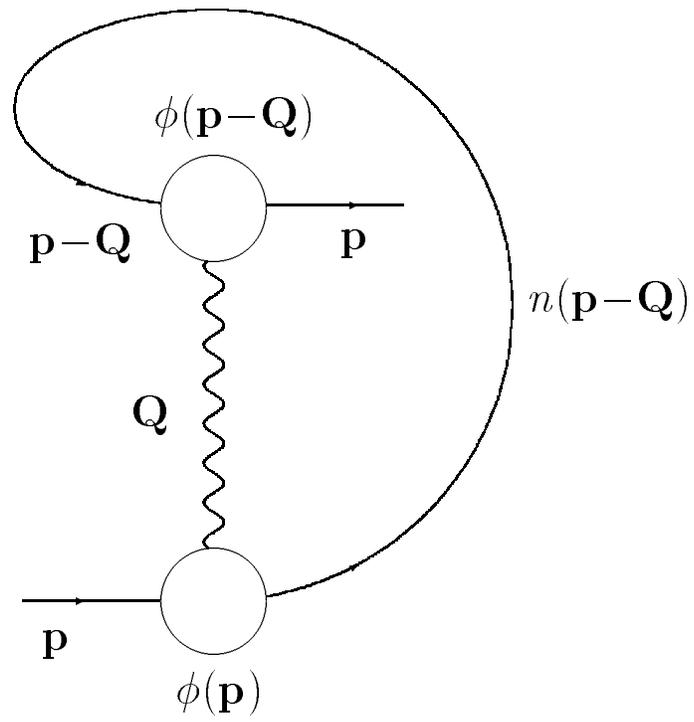
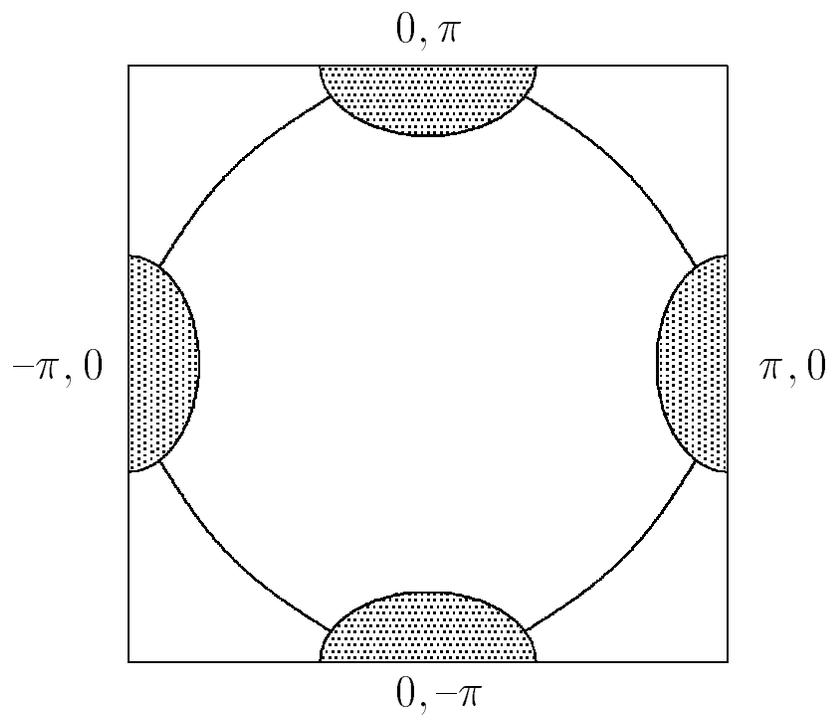
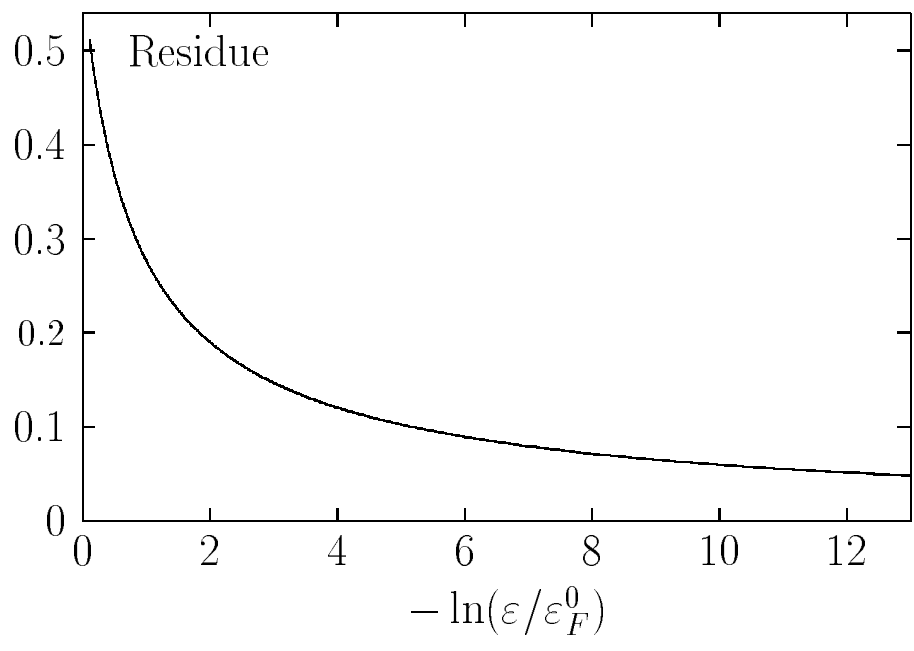
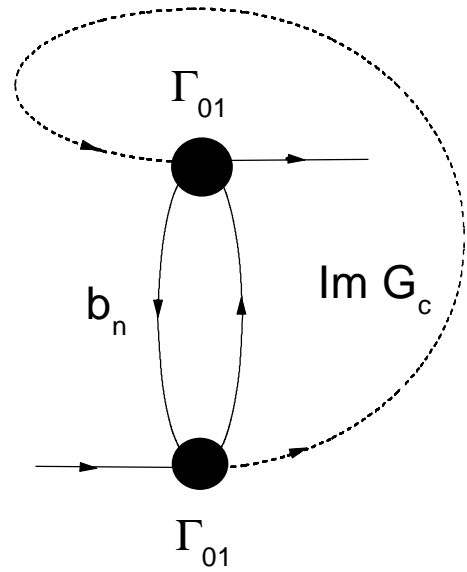
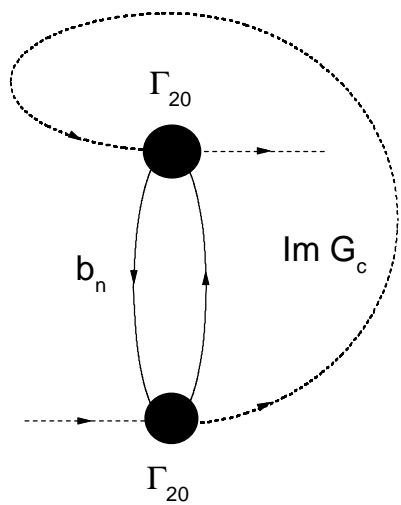
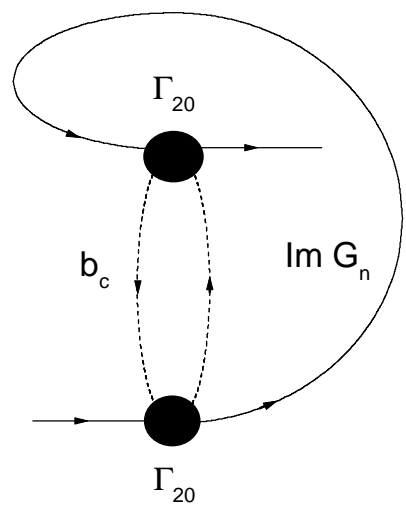
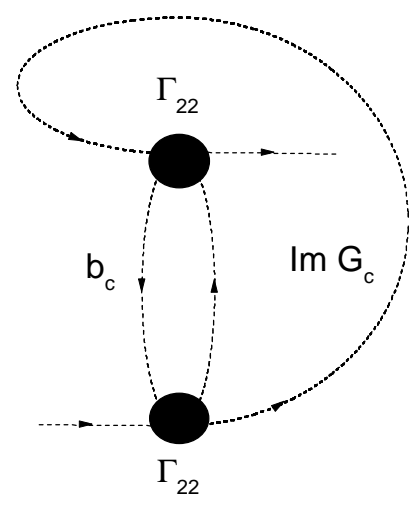
\end{document}